\documentclass[aps,prl,twocolumn,groupedaddress,showpacs]{revtex4}

\usepackage{graphicx,color}
\usepackage{amsmath}

\begin{document}
\title{The nature of the observed free-electron-like state
in a PTCDA monolayer on Ag(111)}

\author{Matthew S. Dyer$^{a}$ and Mats Persson$^{a,b}$}
\affiliation{$^{a}$The Surface Science Research Centre, The University of Liverpool, Liverpool, L69 3BX, UK, \\
$^{b}$Department of Applied Physics, Chalmers University of Technology, SE-412 96 Gothenburg, Sweden}

\begin{abstract}
A free-electron like band has recently been observed in a monolayer of PTCDA (3,4,9,10-perylene tetracarboxylic dianhydride) molecules on Ag(111) by two-photon photoemission [Schwalb et al., Phys. Rev. Lett. {\bf 101}, 146801 (2008)] and scanning tunneling spectroscopy [Temirov et al., Nature {\bf 444}, 350 (2006)]. Using density functional theory calculations, we find that the observed free-electron like band originates from the Shockley surface state band being dramatically shifted up in energy by the interaction with the adsorbed molecules while it acquires also a substantial admixture with a molecular band. 
\end{abstract}


\maketitle
The electronic states at the organic molecule-metal interface are of key importance for electronics and opto-electronics applications since they influence directly the electron transport across the interface.  These interface states are formed from the interaction between highly dispersive states of the metal substrate and weakly dispersive states of the semiconducting organic molecular layers. The molecular states  at the interface shift, broaden and become partially populated or depopulated upon interaction with the metal states. The effect of the organic molecular layers on metal surface states such as the Shockley surface (SS) states on noble metal surfaces is less clear and has recently, attracted a lot of attention from observations of dispersive interface states in these systems.

In many cases upon adsorption of organic molecules on gold and copper surfaces~\cite{Nicetal06,Ziretal09,Scheyetal09,Tametal08,Kanetal07,DyePer08}  the SS states were observed from scanning tunneling spectroscopy (STS) and angle-resolved photoemission to experience small energy shifts. In some cases the fate of the SS and the origin of highly dispersive states turned out to be more complex. A dispersive hybrid band was identified by STS at the interface between a monolayer of charge-transfer complex on a gold surface and was shown by density functional theory calculations to arise from a mixing of metal and molecular states~\cite{Gonetal08}.  A free-electron like band was identified by scanning tunneling spectroscopy in a monolayer of PTCDA (3,4,9,10-perylene tetracarboxylic dianhydride) molecules on the Ag(111) surface and was attributed to originate from the lowest unoccupied molecular orbital (LUMO) of the PTCDA molecule~\cite{TemSouLuiTau06}. This free-electron like band was recently also identified in two-photon photoemission experiments by Schwalb and coworkers~\cite{Schetal08} and they challenged the interpretation proposed by Temirov and coworkers\cite{TemSouLuiTau06}. Based on the observed short life time of electrons in this band, Schwalb and coworkers argued that this band corresponded to an upward shifted SS band by the interaction with the molecular monolayer. Thus there is a need to carry out electronic structure calculations to resolve this outstanding issue about the origin of the observed free-electron like state. 

In this letter we present density functional theory (DFT) calculations of the electronic structure of the PTCDA monolayer on a Ag(111) surface and show that the origin of the observed free-electron-like state is a Shockley surface state (SS) that has been shifted up in energy from the interaction with the molecular monolayer. We also find that the observed free-electron state has acquired character from the LUMO+1 of the adsorbed molecule.

The electronic and geometric structure of the molecular monolayer on the metal surface were determined from periodic density functional theory (DFT) calculations using the projector augmented wave method~\cite{KreFur96}, as implemented in the plane-wave based VASP code~\cite{KreJou99}.  The surface unit cell of the two structurally inequivalent PTCDA molecules adsorbed on a Ag(111) slab is described in detail by Rohlfing and coworkers~\cite{RohTemTau07}. The vacuum region is about 16.5 {\AA}. The full geometry relaxation of the molecular monolayer and the two outermost layers of the Ag slab was carried out using a four layer slab with a plane wave cutoff of 400 eV on a 6$\times$4$\times$1 $k$-point grid until the forces acting on the ions were smaller in magnitude than 0.02 eV/{\AA}. In order to have a good description of the Shockley surface band, a 12 layer slab was used in the calculation of the electronic states. This slab was obtained by adding 8 layers of Ag atoms at their bulk positions to the bottom of the structurally optimized four layer slab

Following the widely used Tersoff--Hamann approximation~\cite{TerHam83}, the differential conductance measured in STS experiments was approximated using the local density of states (LDOS) calculated for the 12 layer slab at the position of the tip apex. The molecular orbital projected density of states (MO-PDOS) was calculated by projecting the wave functions of the full system onto the wave functions of the isolated monolayer kept in the same geometric structure as for the adsorbed monolayer. The wave vector resolved LDOS and PDOS were calculated along a line from the $\Gamma$ point to the surface Brillouin zone (SBZ).

Supported by the findings of Rohlfing and coworkers~\cite{RohTemTau07,RohBre08}, we used the local density approximation (LDA) for the exchange--correlation functional. The generalized-gradient approximation (GGA) was found to give a too large bonding distance and a too weak bonding. A recent study by Romaner and coworkers~\cite{Rometal09} showed that an inclusion of non-local interactions using a recently developed van der Waals functional~\cite{DioRydSchLanLun04} in the DFT calculations improved the binding energy but gave also a too large bonding distance. In contrast the LDA was found to give a bonding geometry in much better agreement with standing X-ray diffraction data~\cite{Hauetal05,HauKaretal05}. The adsorption energies versus vertical distance obtained in the LDA were shown to be in good agreement with the results based on a more fundamental approach to weak adsorption.  In this approach the exact exchange energy is combined with a random phase approximation for the correlation energy as obtained from the adiabatic connection dissipation fluctuation theorem~\cite{RohBre08}. Our calculated bonding distances are in good agreement with the results obtained by Rohlfing and coworkers~\cite{RohTemTau07}. 

\begin{figure}[h]
\includegraphics[width=8.5cm]{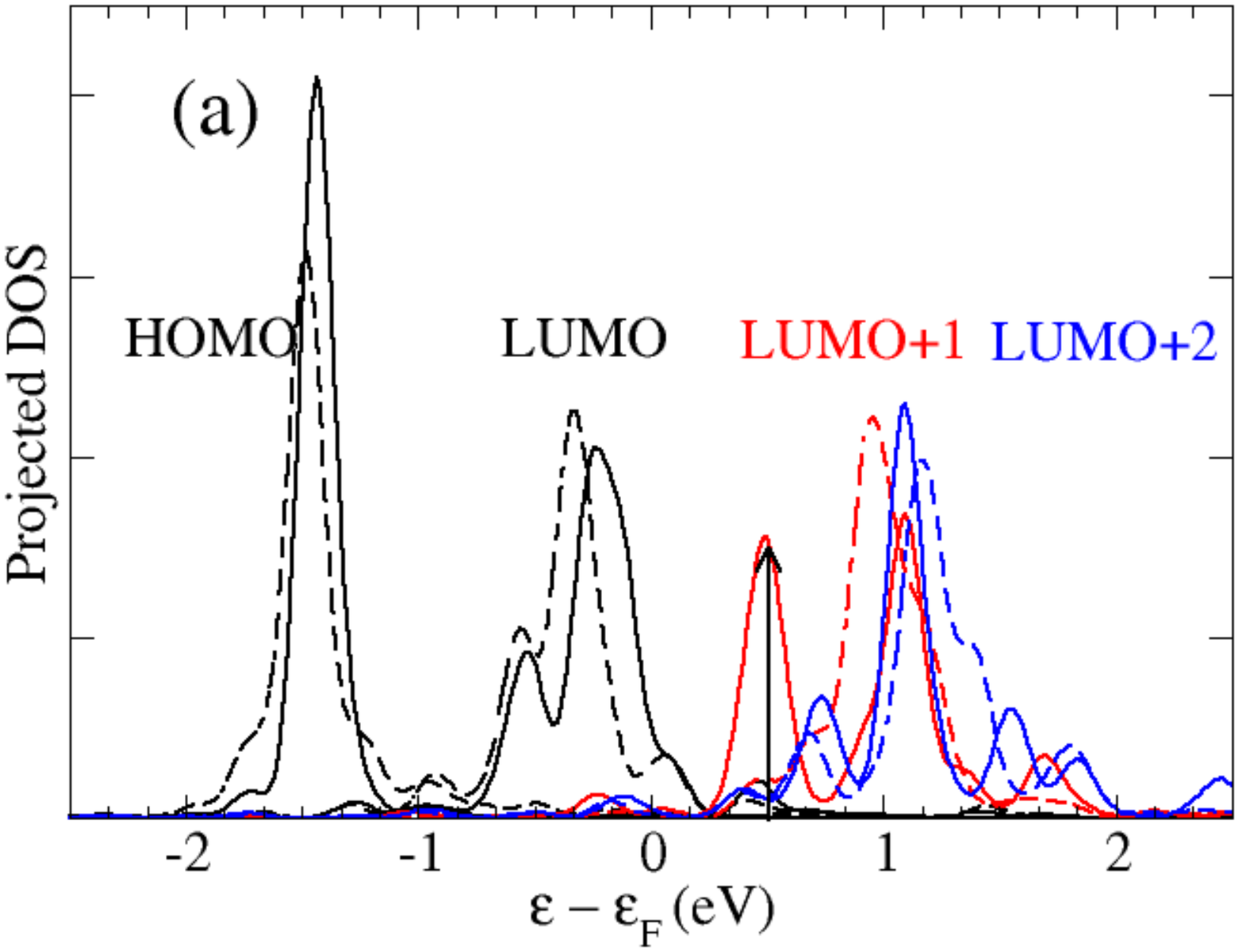} 
\includegraphics[width=8.5cm]{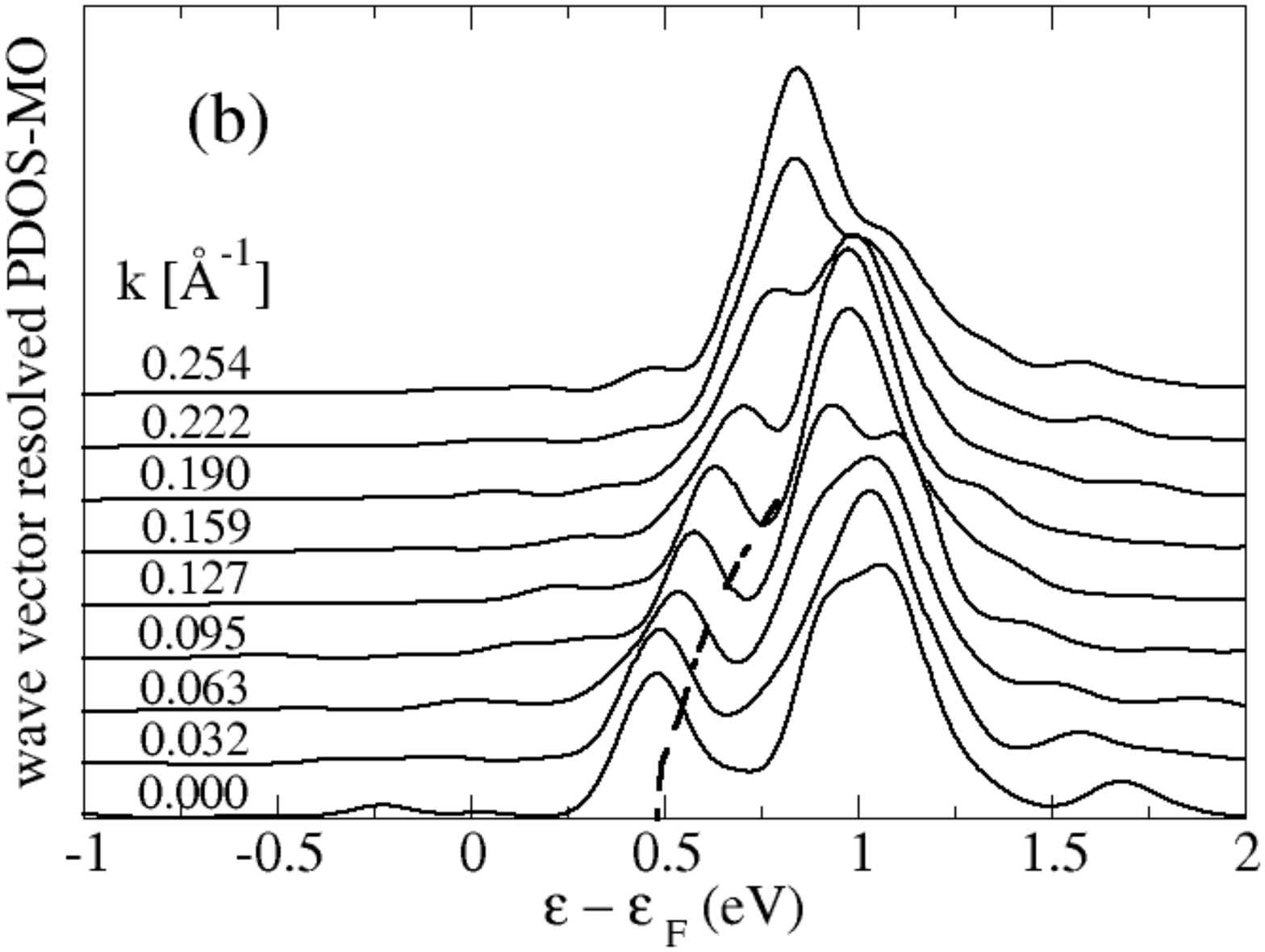}
\caption{Projected DOS (PDOS) of the adsorbed PTCDA monolayer on the frontier molecular orbitals of the isolated PTCDA monolayer. In (a) PDOS on the HOMO, LUMO, LUMO+1, and LUMO+2 are taken at the $\Gamma$ point in the SBZ. The dashed and solid lines correspond to the projected DOS of the two nearly degenerate MOs of the two molecules in the surface unit cell. Their energies when aligning the vacuum levels of the isolated and the adsorbed monolayer are indicated by vertical arrows. The dispersion of the projected DOS on the LUMO+1 is shown in (b) where thick the dashed line indicate the energies of the peak positions in the PDOS. Both LUMO+1 states are included in (b). 
\label{fig:PDOSMO}}
\end{figure}

The characters of the electronic states of the adsorbed molecular monolayer on the Ag surface are revealed by the calculated projected DOS on the frontier orbitals of the isolated monolayer, shown in Fig.~\ref{fig:PDOSMO}(a) at the $\Gamma$ point in the SBZ. For the isolated monolayer, the two molecules in the surface unit cell result in two nearly degenerate bands for each highest occupied molecular orbital (HOMO),  lowest (LUMO), second lowest (LUMO+1) and the third lowest (LUMO+2) unoccupied molecular orbitals.  The energy splittings are only about 0.1 eV at the $\Gamma$ point in the SBZ and demonstrate that the direct interactions between the molecules are small.  Note that the LUMO+1 and LUMO+2 are nearly degenerate and their ordering is reversed and hence their labeling compared to the one obtained by Rohlfing and coworkers~\cite{RohTemTau07}.
 
Upon adsorption, the frontier orbitals shift down in energy and are broadened by the interaction with the metal states. The HOMOs are largely unaffected and shifted down in energy with about 0.41 eV.  The two LUMOs broaden and become more or less fully occupied with peak positions -0.32 and  -0.21 eV being close to the measured peak positions of the P1 state by STS~\cite{TemSouLuiTau06}. This assignment is in agreement with previous  studies~\cite{RohTemTau07}. Both the LUMO+1s and LUMO+2s broaden appreciable and develop satellite structures but remain unoccupied. Of particular note is the pronounced satellite peak at the energy of about 0.5 eV in the PDOS on the LUMO+1s. 

The energy of this satellite peak in the PDOS at the $\Gamma$ point is close to the onset of about 0.6 eV and 0.7eV of the free-electron-like band (P2 state) observed in 2PPE~\cite{Schetal08} and STS~\cite{TemSouLuiTau06}, respectively. Furthermore, this peak has a free-electron like dispersion away from the $\Gamma$ point as shown in Fig.~\ref{fig:PDOSMO}(b) by the wave vector resolved PDOS on the two LUMO+1s in the SBZ. The calculated effective mass of 0.5 $m_e$  is close to the effective mass of about 0.39 $m_e$ and 0.47 $m_e$ extracted from the observed FE band in 2PPE~\cite{Schetal08} and STS~\cite{TemSouLuiTau06}, respectively.  This close agreement between observed and calculated onsets and effective masses suggests that the calculated FE band corresponds to the observed FE band. Note that in this case of relatively delocalized states, the Kohn-Sham states should provide a reasonable zero-order approximation of the one-electron addition spectrum probed by STS. Further support for this assignment in the case of STS comes from a comparison of calculated LDOS with observed STS spectra and images.

\begin{figure}[h]
\includegraphics[width=8.5cm]{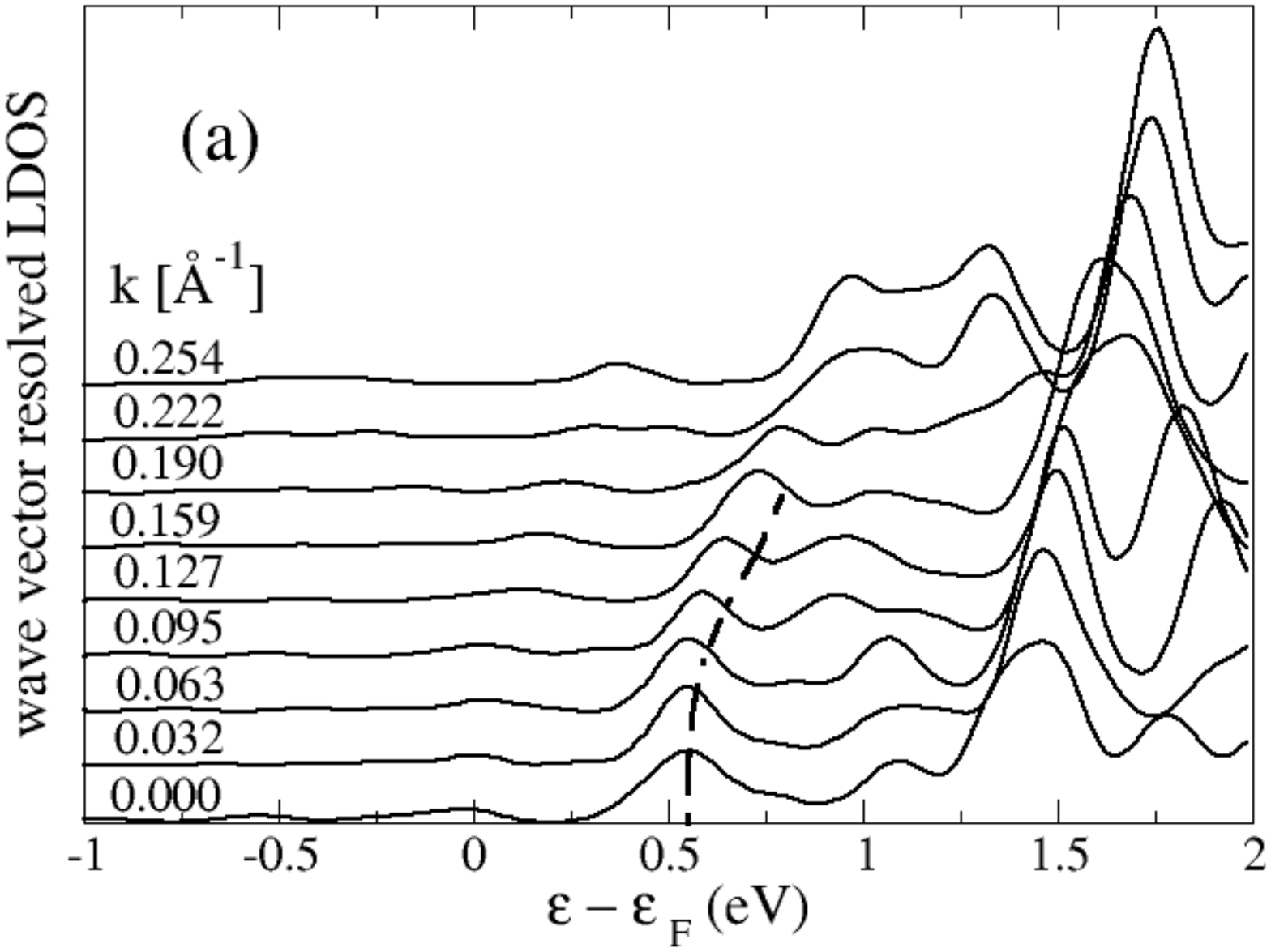}
\includegraphics[width=8.5cm]{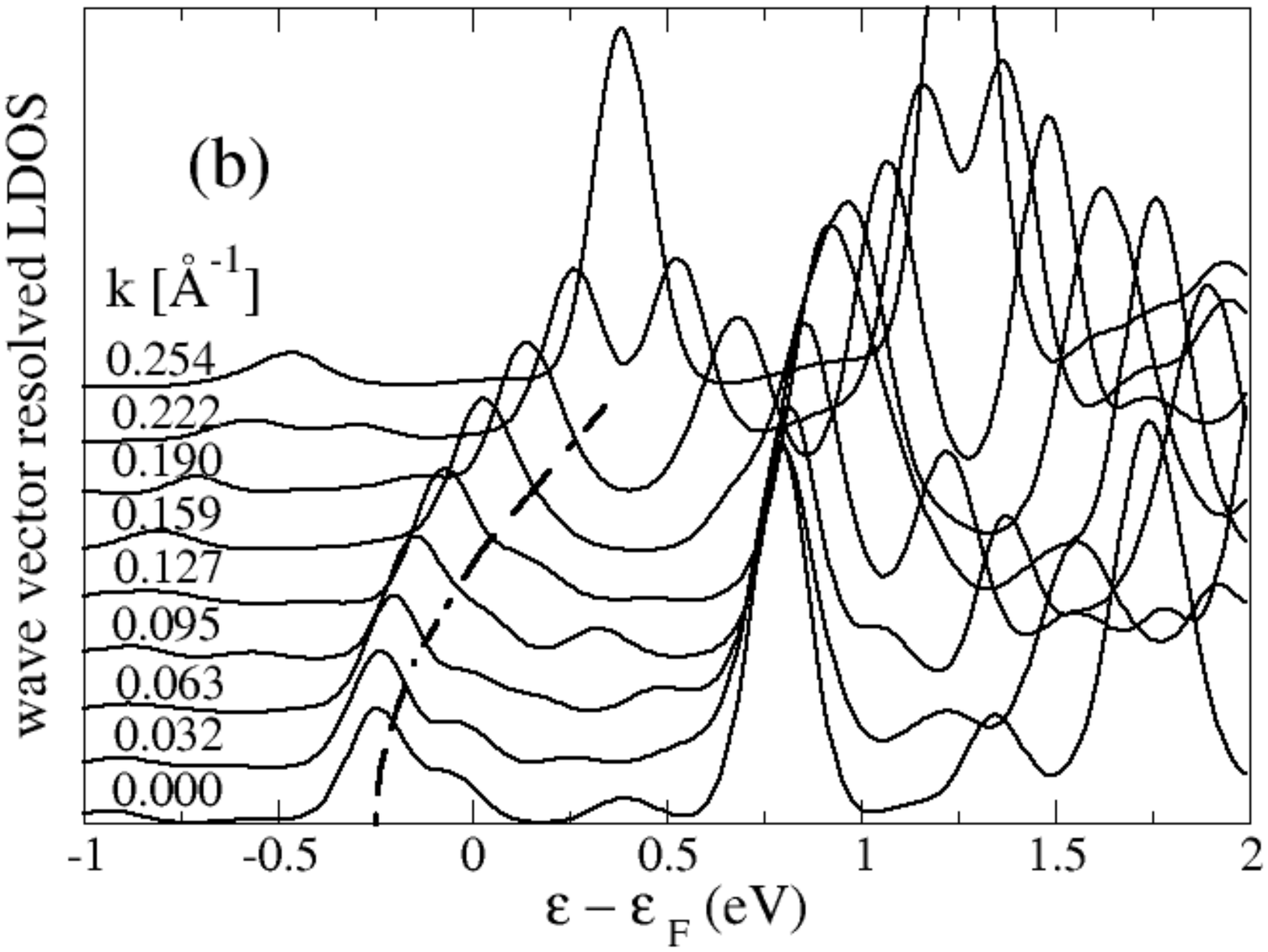}
\caption{Wave vector resolved LDOS of the PTCDA monolayer adsorbed on the Ag slab (a) and the bare side of the Ag slab (b) at distance of 7 {\AA} from the top Ag layer. The magnitude $k$ of the parallel momentum along the direction in the surface Brillouin zone boundary is indicated and $k$ = 0.254 \ {\AA}$^{-1}$ corresponds to the zone boundary. The thick dashed lines indicate the energies of the peak positions in the LDOS. The states were broadened by 0.1 eV. 
\label{fig:WaveLDOS}}
\end{figure}

\begin{figure}[h]
\includegraphics[width=4.25cm]{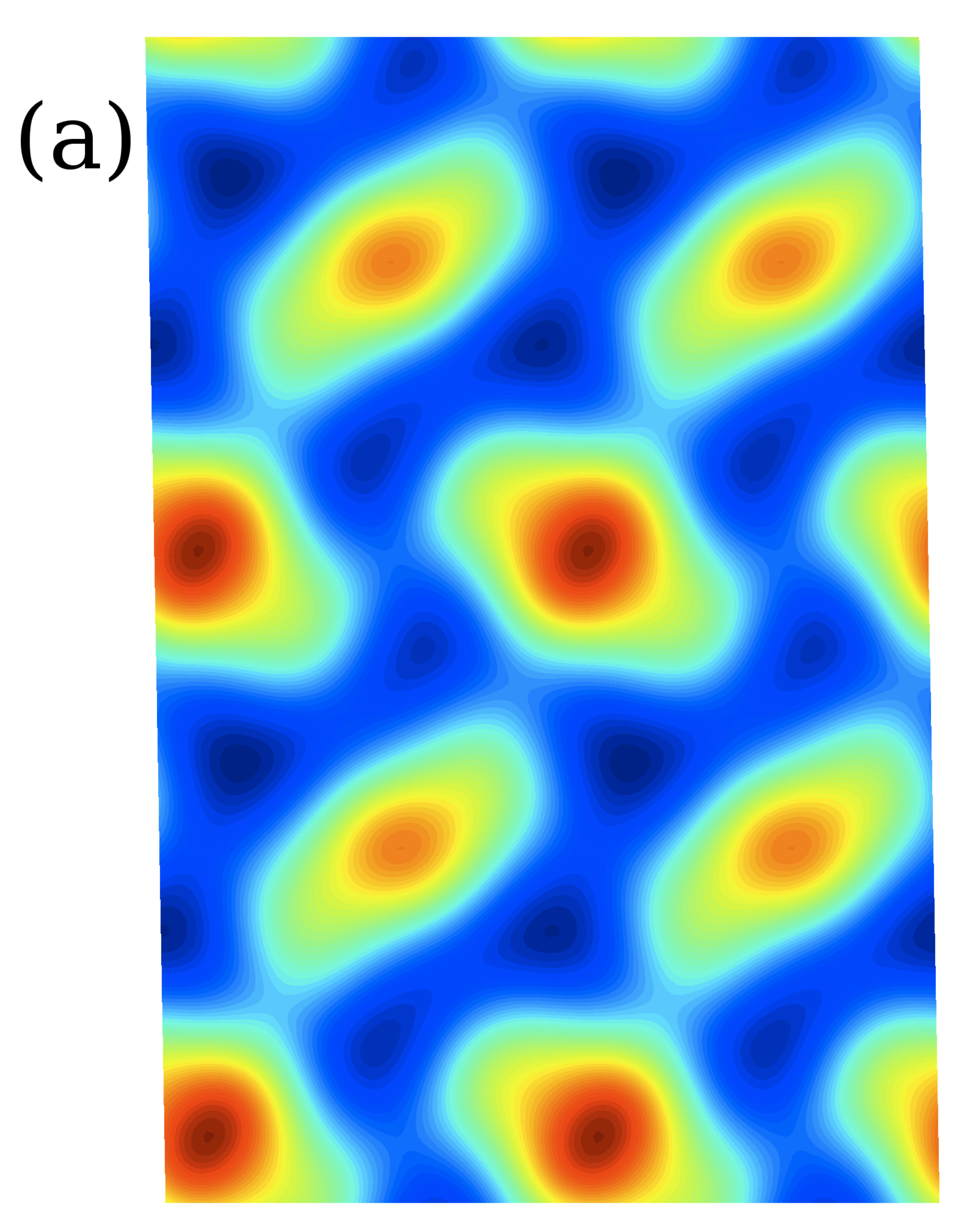} \includegraphics[width=4.25cm]{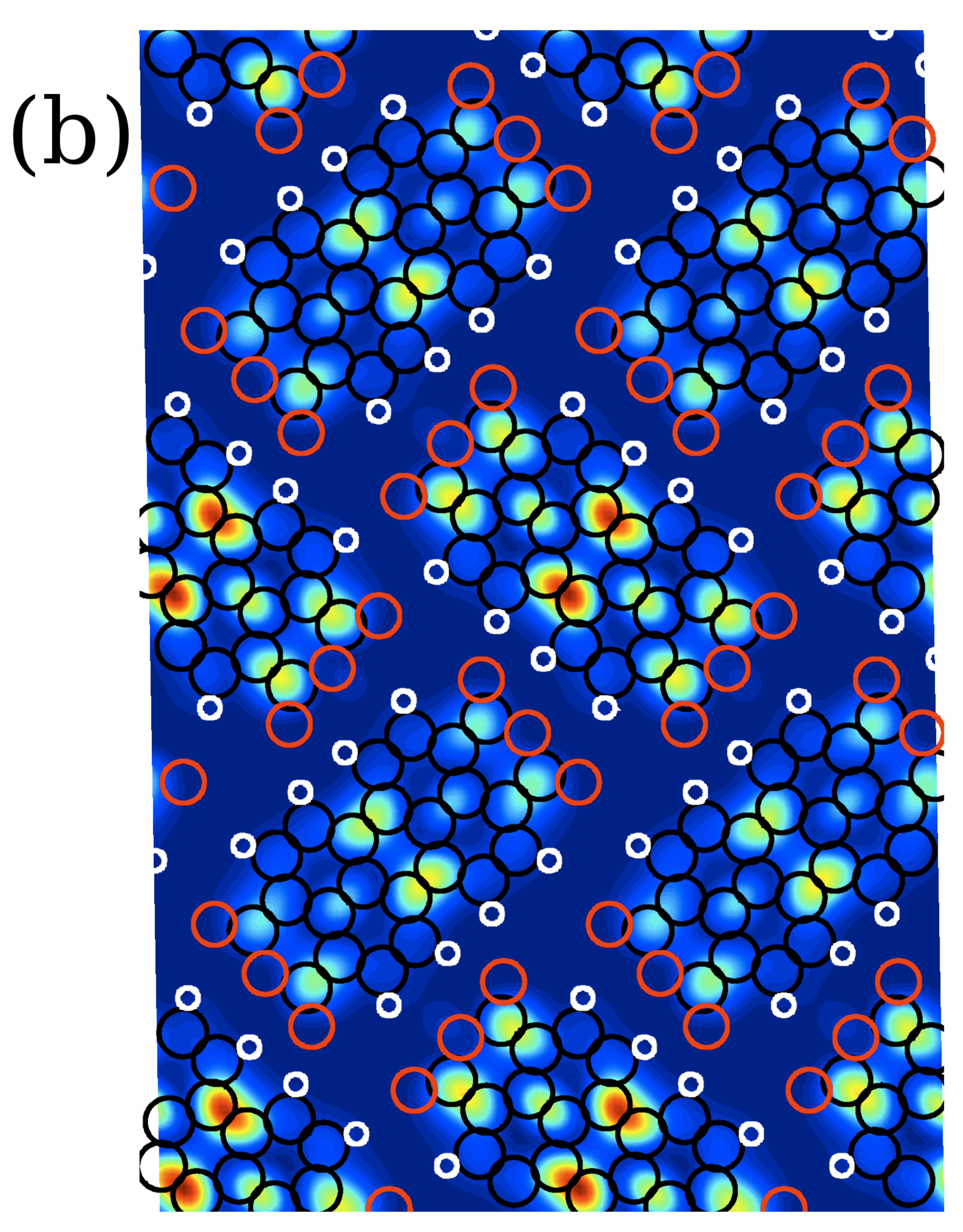}
\caption{Contour plots of calculated LDOS images of the PTCDA monolayer on Ag(111) for vertical tip-molecule distances of about (a) 7.2 {\AA} and (b) 1.3 {\AA} at an energy 0.50 eV above the Fermi level and at the $\Gamma$ point in the SBZ. Red and blue contours correspond to high and low LDOS, respectively. In (b) the H, C and O atoms are represented by white, black and red circles, respectively.
\label{fig:LDOSImage}}
\end{figure}

As shown in Fig.~\ref{fig:WaveLDOS}(a), the FE band identified in the PDOS on the LUMO+1 shows also up in the wave vector resolved LDOS  and demonstrates that it should be observable in STS. The assignment of the observed FE band to the calculated FE band is further corroborated by the calculated LDOS image at the onset of the FE band corresponding to an energy of 0.50 eV and the $\Gamma$ point in the SBZ. This image is shown in Fig.~\ref{fig:LDOSImage}(a) and is in good agreement with the observed STS image (Inset in Fig. 3(a) of Ref.~\onlinecite{TemSouLuiTau06}). At this tip apex-surface distance this image shows no nodal structure and does not resemble  any MO of the PTCDA molecules. However, at a much closer distance of about 1.3 {\AA} from the molecular monolayer the calculated LDOS image in Fig.~\ref{fig:LDOSImage}(b) has a similar nodal structure to the orbital density of the LUMO+1 state as expected from the strong overlap of the FE state with the LUMO+1 state in the PDOS (Fig.~\ref{fig:PDOSMO}(a)). Only LUMO+1 of the isolated molecule is totally symmetric among the frontier orbitals (see LUMO+2 in Fig. 3 of Ref.~\cite{RohTemTau07}) and can interact with the FE band at the $\Gamma$ point.

On the bare side of the slab, the Shockley surface state (SS) band is clearly identified in the wave vector resolved LDOS (Fig.~\ref{fig:WaveLDOS}(b)). In this calculation the onset of this SS band is somewhat below the observed value of about -0.06 eV\cite{Reietal01} and the calculated effective mass of about 0.36 $m_e$ is close to the measured value of about 0.40 $m_e$~\cite{Reietal01}.

\begin{figure}[h]
\includegraphics[width=7cm]{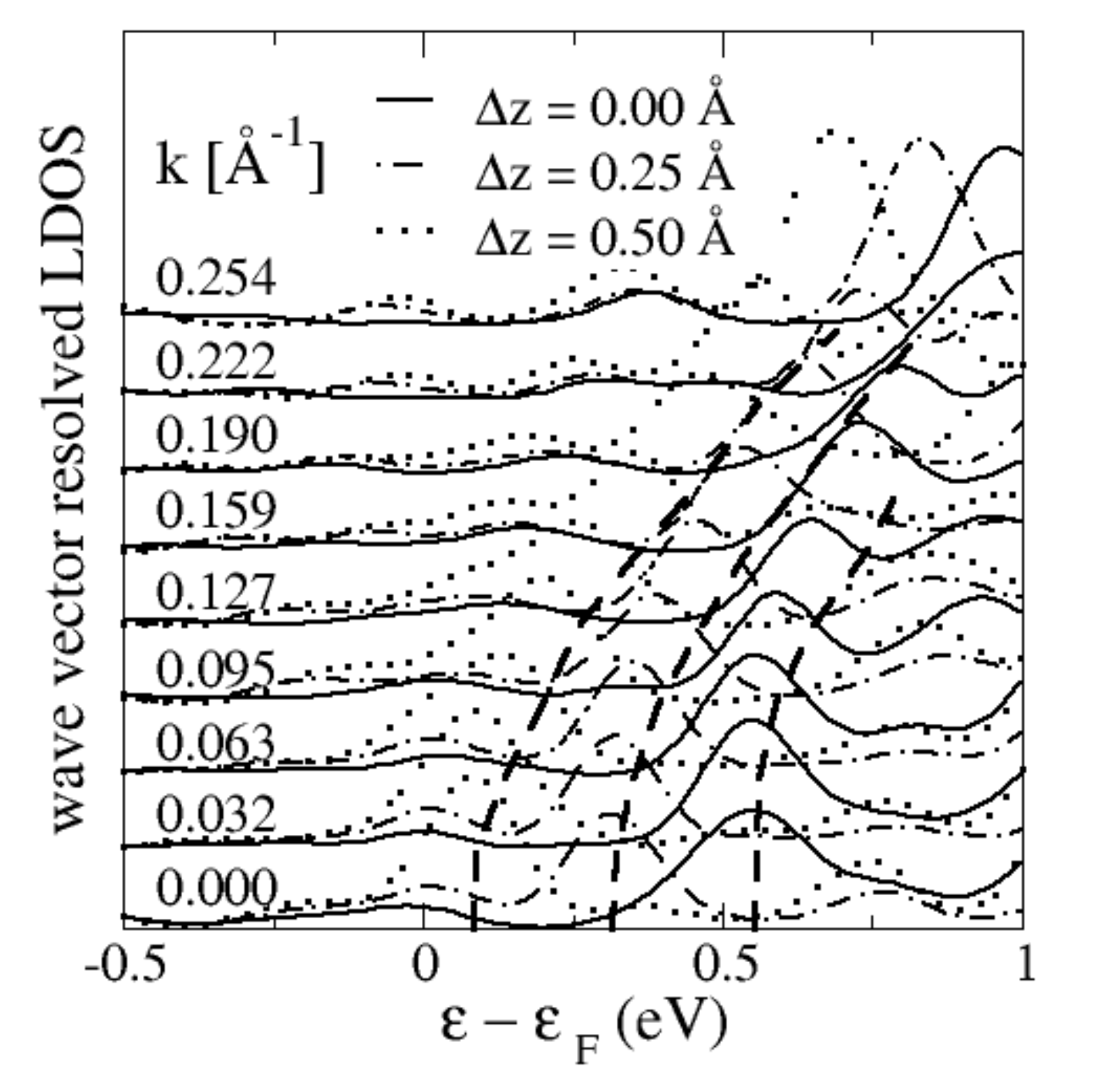}
\caption{Evolution of wave vector resolved LDOS of the PTCDA monolayer adsorbed on the Ag slab with the height $\Delta z$ of the monolayer from its equilibrium position on the surface. The magnitude $k$ of the parallel momentum along the direction in the surface Brillouin zone boundary is indicated and $k$ =0.254 \ {\AA}$^{-1}$ corresponds to the zone boundary. The thick dashed lines indicate the energies of the peak positions in the LDOS. The states were broadened by  0.1 eV. 
\label{fig:WaveLDOSDist}}
\end{figure}

The strong overlap of the FE band with the LUMO+1 (Figs.~\ref{fig:WaveLDOS} and \ref{fig:LDOSImage}), suggests that LUMO+1 is directly involved in the formation of the FE  band as proposed by Temirov and coworkers~\cite{TemSouLuiTau06}. The alternative proposal by Schwalb and coworkers~\cite{Schetal08} that the FE band originates from the SS band of the bare surface has been investigated by calculating the evolution of the FE band with the vertical height of the molecular monolayer from the surface. In Fig.~\ref{fig:WaveLDOSDist}, we depict the wave vector resolved LDOS as a function of a rigid outward shift $\Delta z$ of the molecular monolayer from its vertical equilibrium distance. The FE band shifts rapidly down in energy with increasing $\Delta z$. At $\Delta z$ = 0.5 {\AA}  the onset has already shifted down to about 0.1 eV above the Fermi energy and approaches the onset of the SS. Furthermore, the dispersion increases somewhat with increasing $\Delta z$ and becomes closer to the dispersion of the SS band. Thus the evolution of the FE band with the height of the molecular monolayer shows that the FE band corresponds to a SS band shifted upwards in energy by the interaction of the Ag surface with the adsorbed monolayer but acquire an admixture with the LUMO+1.

In conclusion, we have shown from a density functional theory study that the observed free-electron band in scanning tunneling spectroscopy and in two photon photoemission spectroscopy of an adsorbed monolayer of PTCDA molecules on a Ag(111) surface originates from the Shockley surface state being dramatically shifted up in energy by the interaction with the adsorbed molecules. The band acquires also a substantial admixture with the LUMO+1 bands of the molecular monolayer. This finding shows that the metal surface states play in important role in forming dispersive states at the metal organic interface with potentially important implications for the electron transport across the interface.

\begin{acknowledgments}
The authors would like to thank the Marie Curie Research Training Network PRAIRIES, contract MRTN-CT-2006-035810 and the Swedish Research Council (VR) for financial support, and the University of Liverpool  (UoL) and the Swedish National Infrastructure of Computing (SNIC) for allocation of computational resources. MSD is grateful for funding of postdoctoral fellowship by UoL 
\end{acknowledgments}

\bibliography{MPref,MyPapers}

\end{document}